\documentclass[pre,fleqn]{w-artpre}
\usepackage{times}
\usepackage{amsfonts,amsmath,axodraw}

\newcommand{\Supertwistor}{\Cset \mathrm{P}^{3|4}}
\newcommand{\dbar}{\bar{\partial}}
\newcommand{\Dbar}{\overline{D}}
\newcommand{\Tr}{\mathrm{Tr}}
\newcommand{\tlambda}{\widetilde{\lambda}}
\newcommand{\tchi}{\widetilde{\chi}}
\newcommand{\tphi}{\widetilde{\phi}}
\newcommand{\Dirac}{D\!\!\!\!\slash}

\newcommand{\Cset}{{\,\,{{{^{_{\pmb{\mid}}}}\kern-.47em{\mathrm C}}}}}

\newcommand{\p}{\partial}

\newcommand{\half}{\frac{1}{2}}
\newcommand{\diff}{\mathrm{d}}

\newcommand{\ra}{\rightarrow}

\newcommand{\grl}{\lambda}
\newcommand{\gre}{\epsilon}

\newcommand{\Acal}{{\mathcal A}}

\newcommand{\Ncal}{{\mathcal N}}
\newcommand{\Lcal}{{\mathcal L}}
\newcommand{\Scal}{{\mathcal S}}
\newcommand{\Wcal}{{\mathcal W}}
\newcommand{\Bcal}{\mathcal{B}}

\newcommand{\Vcal}{\mathcal{V}}
\newcommand{\Ocal}{\mathcal{O}}

\newcommand{\Urm}{{\mathrm U}}

\newcommand{\SU}{\mathrm{SU}}

\newcommand{\ip}[1]{\langle{#1}\rangle}
\newcommand{\hb}{\overline{h}}

\newcommand{\preprint}{QMUL-PH-05-20}

\begin{document}

\author[M.Kulaxizi]{Manuela Kulaxizi\inst{1,}\footnote{Email: kulaxizi@insti.physics.sunysb.edu}}
\author[K.Zoubos]{Konstantinos Zoubos\inst{2,}\footnote{Email: k.zoubos@qmul.ac.uk . 
Talk presented at the RTN meeting ``Constituents, Fundamental
Forces and Symmetries of the Universe'', Corfu, Greece, Sept. 20-26, 2005.}}
\address[\inst{1,}]{ C. N. Yang Institute for Theoretical Physics,
State University of New York,\\ Stony Brook, New York 11794-3840, U.S.A.}
\address[\inst{2,}]{Queen Mary, University of London, Mile End Road,
London E1 4NS, U.K.}

\title{Marginal Deformations of Tree--Level {$\mathcal{N}=4$} SYM from Twistor 
String Theory}

\begin{abstract}
The topological B--model with target the supertwistor space $\Supertwistor$
is known to describe perturbative amplitudes of $\Ncal=4$ Super Yang--Mills
theory. We review the extension of this correspondence to the superconformal gauge
theories that arise as marginal deformations of $\Ncal=4$ by considering
the effects of turning on a certain closed string background, which 
results in non--anticommutativity in the fermionic directions of 
$\Supertwistor$. We generalise the twistor string prescription for
amplitudes to this case and illustrate it with some simple examples. 
\end{abstract}

\maketitle

Witten's original formulation of twistor string theory \cite{Witten0312}
relates the perturbative
expansion of $\Ncal=4$ Super Yang--Mills theory to the $D$-instanton 
expansion 
of the topological B-model on supertwistor space $\Supertwistor$. 
The motivation for this reformulation was the fact that certain 
Yang--Mills amplitudes are much
simpler than one would expect from the properties of the individual 
Feynman diagrams they are composed of. Witten's proposal led to the 
development of very efficient calculational tools for scattering 
amplitudes in gauge theories, some aspects of which are summarised in the
review papers \cite{Khoze04,CachazoSvrcek05}.

The starting point for the gauge theory/twistor string correspondence is the fact
that gauge theory tree amplitudes take a particularly simple form
when written in a basis where the helicities of the external particles are 
fixed. For instance, the maximally helicity violating (MHV)
$n$--point amplitude with  $n-2$ positive 
helicity and 2 negative helicity gluons is proportional to the subamplitude
\begin{equation} \label{MHV}
\Acal_{(n)}=\frac{\ip{n-1,n}^4}{\ip{12}\ip{23}\ldots\ip{n1}}
\end{equation}
where the negative helicity gluons are labeled by $n-1$ and $n$.
Here we have decomposed the momenta of the incoming particles in terms
of commuting spinors $\grl,\tlambda$ as $p_{\alpha\dot{\alpha}}=\grl_\alpha
\tlambda_{\dot{\alpha}}$ and defined the holomorphic inner product 
$\ip{ij}=\gre_{\alpha\beta}\grl^\alpha_i\grl^\beta_j$. To recover the
full amplitude for this process we have to add a colour trace for
each cyclic ordering and sum over all subamplitudes, and also incorporate 
a momentum conservation delta function which we have suppressed.

The crucial property of (\ref{MHV}) is the holomorphic dependence on the 
spinors $\grl_i$. As discussed in \cite{Witten0312}, this implies
that MHV amplitudes are supported on genus zero, degree one curves on 
\emph{twistor space}, which is a copy of $\Cset \mathrm{P}^3$ defined by 
the homogeneous coordinates $(\grl^\alpha,\mu^{\dot{\alpha}})$, where the $\mu^{\dot{\alpha}}$ are
related to $\tlambda_{\dot{\alpha}}$ as  
\begin{equation}
\tlambda_{\dot{\alpha}}\ra i\frac{\p}{\p \mu^{\dot{\alpha}}}, \quad
-i \frac{\p}{\p\tlambda^{\dot{\alpha}}}\ra \mu_{\dot{\alpha}}.
\end{equation} 
 Non--MHV amplitudes with $q$ negative helicity gluons
were similarly shown to be supported on curves of degree $q-1$ in twistor
space.

In \cite{Witten0312} this fact was combined with the observation that if one
adds four fermionic coordinates $\psi^A$ (and their conjugates) 
to $\Cset \mathrm{P}^3$, the resulting
supermanifold $\Supertwistor$ is Calabi--Yau, in the sense of admitting
a globally defined holomorphic volume form. It is thus a suitable target
space for the topological string theory known as the B--model, whose
open string field theory action is given by holomorphic Chern--Simons 
theory
\begin{equation} \label{SHCS}
\Scal=\frac{1}{2}\int_{D5}\Omega\wedge
\Tr\left(\Acal\dbar\Acal+\frac{2}{3}\Acal\wedge\Acal\wedge\Acal\right)\;
\end{equation}
and the model is actually not defined on the whole of $\Supertwistor$, 
but on a ``$D5$--brane'' which fills the whole of the bosonic part
but lies on the locus $\bar{\psi}^{\bar{A}}=0$. Thus the $\Acal$ are 
superfields which depend only on the holomorphic coordinates $\psi^A$, 
and can be expanded in components as:
\begin{equation} \label{Superfield}
\begin{split}
\Acal=A+\xi\lambda&+\psi^I\chi_I+\xi\psi^I\phi_I
+\half\psi^I\psi^J\gre_{IJK}\tphi^K\\&+\half\xi\psi^I\psi^J\gre_{IJK}\tchi^K
+\frac{1}{3!}\psi^I\psi^J\psi^K\gre_{IJK}\tlambda
+\frac{1}{3!}\xi\psi^I\psi^J\psi^K\gre_{IJK}G\;.
\end{split}
\end{equation}
In this expansion we have split the fermionic coordinates as $\psi^0=\xi,
\psi^I=\psi^A$ for $A=1,2,3$ so that we make evident only an $\SU(3)\times\Urm(1)$
out of the $\SU(4)$ symmetry. This will prove convenient for the discussion of 
$\Ncal=1$ theories later on. 
The component fields $(A,\grl,\chi_I,\phi_I,\tphi^I,\tchi^K,\tlambda,G)$ 
live on
$\Cset\mathrm{P}^3$ and are mapped through the Penrose transform to 
four dimensional fields of helicities $(1,\half,\half,0,0,-\half,-\half,-1)$,
which we can combine into a vector multiplet $(A,\lambda,\tlambda,G)$ and
three chiral multiplets $\Phi_I=(\chi_I,\phi_I)$ (and their conjugates
$\tilde{\Phi}^I=(\tphi^I,\tchi^I)$), which is 
the field content of $N=4$ SYM. Having established the equivalence of
the spectrum, we can now look at the interactions encoded in the action
(\ref{SHCS}) and find that after transforming to four dimensions they actually 
correspond to those of \emph{self--dual} $\Ncal=4$ SYM rather than of the full theory.

So at first it seems that the B--model can reproduce only the self--dual part
of Yang--Mills amplitudes. However, building on work by Nair \cite{Nair88}, Witten 
showed that
the full $\Ncal=4$ amplitudes do arise, but as nonperturbative effects, 
through $D1$--instantons \cite{Witten0312}. For the MHV case, these wrap holomorphic 
degree one, genus zero curves in $\Supertwistor$ whose embedding is given by
\begin{equation} \label{Embedding}
\mu_{\dot{\alpha}}+x_{\alpha\dot{\alpha}}\grl^\alpha=0\;\; \text{and}\;\; 
\psi^A+\theta^{A}_\alpha\grl^\alpha=0\;.
\end{equation}
Thus the curves in twistor space on which the amplitudes are supported are now 
seen to correspond to $D1$--instantons of the B--model on $\Supertwistor$.  
To get a covariant answer we have to integrate over the moduli space
given by the choices of $(x,\theta)$. We are led to the following
prescription for MHV amplitudes:
\begin{equation} \label{Amplitudeformula}
\Acal_{(n)}=\int\diff^4 x\diff^8\theta\, w_1\cdot w_2\cdots w_n 
\ip{J_1\cdot J_2\cdots J_n}
\end{equation}
where the $J$s are currents living on the brane worldvolume whose
OPEs will lead to the denominator of the MHV amplitude and the 
$w_i$s are the coefficients of each field in the superfield expansion
(\ref{Superfield}) and will give the momenta in the numerator of 
(\ref{MHV}). To integrate over the fermionic coordinates one uses (\ref{Embedding}) to 
express ($\xi, \psi^I$) in terms of the $\theta$s.

In trying to go beyond the case of $\Ncal=4$ SYM, it is natural to look for an
extension to cases with less supersymmetry, or that lack conformal invariance. 
It was quickly understood
that the method of Cachazo, Svr\v{c}ek and Witten \cite{Cachazoetal0403}
for calculating non--MHV diagrams based on MHV vertices
applies not just to $\Ncal=4$ but to a far more general class of theories.
Although the methods that have since been developed for these computations
are extremely efficient (see e.g. \cite{CachazoSvrcek05} for a review and references),
in the process their relationship to string theory has become somewhat 
more vague. This is especially evident in the extension to loop amplitudes,
where although it is relatively straightforward to find a CSW--type prescription on 
the field theory side \cite{Brandhuberetal0407},
so far a clean prescription from the string theory side has
not been found. Thus it is important to search for field theories that
can be described by twistor strings at the same level of detail as the $\Ncal=4$ theory.

At this point it is useful to recall that $\Ncal=4$ SYM is actually just a line in a dim$_C=3$ moduli
space of finite, $\Ncal=1$ conformal field theories. Leigh and Strassler 
\cite{LeighStrassler95}
gave an all--orders proof of the conformal invariance of these theories, 
which can be reached from $\Ncal=4$ SYM through marginal deformations.
The full superpotential is given by
\begin{equation} \label{LSsuperpotential}
\Wcal=i\kappa\mathrm{Tr}\left(e^{i\frac{\beta}{2}}\Phi_1\Phi_2\Phi_3
-e^{-i\frac{\beta}{2}}\Phi_1\Phi_3\Phi_2\right)
+\rho\Tr\left(\Phi_1^3+\Phi_2^3+\Phi_3^3\right)\;.
\end{equation}
The $\Ncal=4$ superpotential $i\mathrm{Tr}\left(\Phi_1[\Phi_2,\Phi_3]\right)$
is recovered by setting $\kappa=1,\beta=\rho=0$. 
Expanding (\ref{LSsuperpotential}) to first order in $\beta$ and $\rho$ 
(it can be shown that $\kappa=1$ to this order), we have
\begin{equation} \label{hSuperpotential}
\Wcal=\Wcal_{\Ncal=4}+\frac{1}{3!}h^{IJK}\Tr(\Phi_I\Phi_J\Phi_K)
\end{equation}
where we have been more general and introduced 
the tensor $h^{IJK}$ which is totally symmetric in its indices, and thus 
lies in the $\mathbf{10}$ of $\SU(3)$. 

Classically this superpotential describes marginal deformations of 
the $\Ncal=4$ lagrangian for any value of $h^{IJK}$. However asking for 
\emph{exactly} marginal deformations requires that we make a very 
particular choice of $h^{IJK}$, i.e. we
need to take the nonzero components to be a linear combination of 
\begin{equation}
\begin{split} \label{hexactlymarginal}
(a)&\quad h^{123}={\beta}, \quad \text{and}\\
(b)&\quad h^{111}=h^{222}=h^{333}=\rho\;
\end{split}
\end{equation}
thus recovering (\ref{LSsuperpotential}), expanded to first order.

Given that the Leigh--Strassler deformed theories are conformal and
just as finite as $\Ncal=4$ SYM is, it is natural to ask whether their
perturbative amplitudes might 
also be described by the B-model on some Calabi-Yau manifold. This was
investigated in \cite{KulaxiziZoubos04} and will be reviewed in the following.

To start discussing this problem, we have to rewrite the action derived from the above
superpotential in a form suited to expanding around the self--dual
sector of the theory. This is done by writing the Yang--Mills action in first--order form 
(introducing an antiselfdual two--form G) and performing helicity--dependent 
rescalings of the fields, after which the component lagrangian takes the form:
\small
\begin{equation} \label{LSaction}
\begin{split}
\Lcal=\mathrm{Tr}&\left[GF+ 
\tlambda \Dirac \lambda+
\frac{1}{2}(D\tphi^{I})(D\phi_{I})+\tchi^{I}\Dirac \chi^{I}
+\lambda[\chi_{I},\tphi^{I}]+
\frac{1}{2}\epsilon^{IJK}\chi_{I}[\phi_{J},\chi_{K}]+
\frac{1}{2}h^{IJK}\chi_{I}\{\phi_{J},\chi_{K}\}\right.\\
-&g^{2} \left(\half G^2 + \tlambda[\tchi^{I},\phi_{I}]+
\frac{1}{2}\epsilon_{IJK}\tchi^{I}[\tphi^{J},\tchi^{K}]+
\frac{1}{2}\hb_{IJK}\tchi^{I} \{\tphi^{J},\tchi^{K}\}
+\frac{1}{8}([\tphi^{I} , \phi_{I}])^{2}\right.\\
&\quad+\left.\left.\frac{1}{8}(\epsilon^{QJK} 
[ \phi_{J} , \phi_{K}]+h^{QJK} \{ \phi_{J} , \phi_{K}\})
(\epsilon_{QIL} [ \tphi^{I} , \tphi^{L}]
+\hb_{QIL}\{ \tphi^{I} , \tphi^{L}\})\right)\right]\;.
\end{split}
\end{equation}
\normalsize
In this lagrangian the terms having no coupling constant dependence 
are naturally associated by supersymmetry with the self--dual part part $GF$ of
the Yang--Mills amplitude, while the terms appearing at order $g^2$
complete the non--self dual interactions of the deformed $\Ncal=4$ SYM.
We can now take the coupling constant $g\ra0$, to recover the
action of self--dual $\Ncal=4$ plus an additional term which only
preserves $\Ncal=1$ supersymmetry. We see that it is proportional to
\begin{equation}
\Tr(h^{IJK}\chi_I\{\phi_J,\chi_K\}).
\end{equation}
So if there indeed exists a generalisation of the B--model side to describe the
Leigh--Strassler theories, we would expect that one can similarly
deform the holomorphic Chern--Simons action (\ref{SHCS}) by adding
a term that is related to the above through the Penrose transform.
Roughly, we would expect this six--dimensional term to look like
\begin{equation}
\int_{C \mathrm{P}^3}\Omega\wedge \Tr(h^{IJK}\chi_I\phi_J\chi_K).
\end{equation}
So we learn that we need to look for a deformation of the 
\emph{cubic} term in the B--model open string field theory. Recalling
that the original cubic term comes from a disk diagram with three open
string insertions, an obvious way to deform it is to introduce a 
closed string background leading to an expansion in insertions of
closed string vertex operators in the bulk of the disk. Schematically,
the picture we obtain is given in Fig. 1. 
 
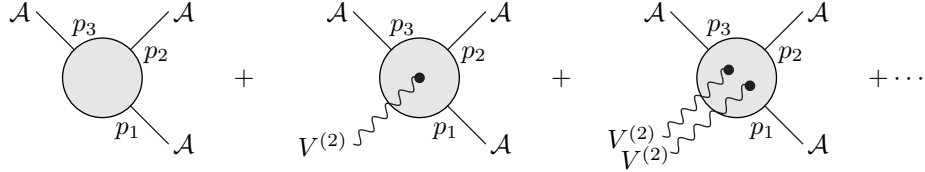
\begin{figure}[h] 
\begin{center}
\begin{picture}(300,80)(0,15)
\put(-20,0){
\Line(25,75)(50,50)\Text(23,75)[r]{$\Acal$} \Text(39,69)[l]{$p_3$}
\Line(75,75)(50,50)\Text(77,75)[l]{$\Acal$} \Text(66,59)[l]{$p_2$}
\Line(75,25)(50,50)\Text(77,25)[l]{$\Acal$} \Text(60,33)[t]{$p_1$}
\GCirc(50,50){15}{.9}}
\put(100,0){\Text(-20,50)[l]{$ +$}
\Line(25,75)(50,50)\Text(23,75)[r]{$\Acal$} \Text(39,69)[l]{$p_3$}
\Line(75,75)(50,50)\Text(77,75)[l]{$\Acal$} \Text(66,59)[l]{$p_2$}
\Line(75,25)(50,50)\Text(77,25)[l]{$\Acal$} \Text(60,33)[t]{$p_1$}
\GCirc(50,50){15}{.9}
\Vertex(50,50){2}
\Photon(25,25)(50,50){-2}{5}\Text(23,25)[r]{$V^{(2)}$}}
\put(220,0){\Text(-20,50)[l]{$+$}
\Line(25,75)(50,50)\Text(23,75)[r]{$\Acal$} \Text(39,69)[l]{$p_3$}
\Line(75,75)(50,50)\Text(77,75)[l]{$\Acal$} \Text(66,59)[l]{$p_2$}
\Line(75,25)(50,50)\Text(77,25)[l]{$\Acal$} \Text(60,33)[t]{$p_1$}
\GCirc(50,50){15}{.9}
\Vertex(47,53){2}\Vertex(55,47){2}
\Photon(22,28)(47,53){-2}{5}\Text(20,28)[r]{$V^{(2)}$}
\Photon(25,22)(55,47){-2}{5}\Text(25,20)[r]{$V^{(2)}$}}
\put(320,0){\Text(0,50)[l]{$+\cdots$}}
\end{picture}
\end{center}
\caption{Closed string contributions to an open string disk diagram at
orders $\Ocal(0),\Ocal(V),\Ocal(V^2)...$.} 
\end{figure}
In Fig. 1 we have fixed the three open string vertex operators
at some positions $p_1,p_2,p_3$ on the boundary of the disk, leaving
the positions of the closed string vertex operators as moduli to
be integrated over the disk. Accordingly, we  have inserted not the closed 
string vertex operator itself, but rather its second--order descendant
$V^{(2)}$ (see \cite{KulaxiziZoubos04} for more details).
However, it turns out that it is easier to calculate the correlation function
at $\Ocal(V)$ by instead fixing the position of the closed string insertion
and integrating over two of the open string insertion positions on the 
boundary, where we now have to insert open string descendants $A^{(1)}$. 
To do this we make use of the equality 
\begin{equation} \label{equivalence}
\ip{\int V^{(2)}\Acal_{p_1}\Acal_{p_2}\Acal_{p_3}}_D
=\ip{V\Acal_{p_1}\int \Acal_{p_2}^{(1)}\int \Acal_{p_3}^{(1)}}_D
\end{equation}
which was shown in \cite{HofmanMa00,Herbstetal04}.

To proceed, we have to be more specific about the background $V$ we would
like to turn on. First of all, BRST invariance strongly restricts the
form of $V$. Furthermore, it is clear that since
spacetime conformal invariance (which we want to preserve) is linked to the 
symmetries of twistor space $\Cset \mathrm{P}^3$, we cannot introduce
a background that deforms this space. To avoid deforming $\Cset\mathrm{P}^3$,
we thus choose $V$ to lie along and depend on only the \emph{fermionic}
directions of $\Supertwistor$. Given also that we want to preserve 
an $\Ncal=1$ out of the original $\Ncal=4$, we split the four fermionic
directions 
in an $\SU(3)\times\Urm(1)$ invariant way as as we did in (\ref{Superfield}), 
and choose $V$ to lie along the 
$\psi^I$. Finally, guided
by the specific form of the marginal deformation we want to reproduce,
we pick a very particular dependence on the $\psi^I$'s, to finally obtain
\begin{equation} \label{VV}
V=\frac{1}{2}\Vcal^{IJ}_{\;\;\;KL}\psi^K\psi^L\vartheta_I\vartheta_J,
\quad \text{where}\quad 
\Vcal^{IJ}_{\;\;\;KL}=h^{IJQ}\epsilon_{QKL}\;.
\end{equation}
and $h^{IJK}$ is the same symmetric $\SU(3)$ tensor that appears in 
(\ref{hSuperpotential}). The $\vartheta_I$ are B-model worldsheet 
fields of ghost number one that can be used to construct physical vertex
operators. They are fermions in the usual case of a bosonic target space,
but in this supermanifold case they can be bosonic if they are along
the odd directions, which is precisely what happens here. Thus we see
why the upper indices in $\Vcal^{IJ}_{\;\;\;KL}$ have to be symmetric. The 
vertex operator $V$ can be shown to be in the BRST cohomology of
the B--model on $\Supertwistor$ and is thus a good operator to deform our
theory with.

Calculating the first--order effect of this background on the
component action, we eventually find \cite{KulaxiziZoubos04}
\begin{equation}\label{Deformedcomponentaction}
\begin{split}
\Scal_{dHCS}=&\int_{\mathbb{C}\mathrm{P}^3}\Omega\wedge
\Tr\left(G\wedge F+\tlambda\wedge\Dbar\grl-\tchi^I\wedge\Dbar\chi_I
+\tphi^{I}\wedge\Dbar\phi_I \right.\\
&\;\left.-\lambda\wedge(\chi_I\wedge\tphi^I+\tphi^I\wedge\chi_I)
+(\gre^{IJK}+h^{IJK})
\chi_I\wedge\phi_J\wedge\chi_K\right)\;.
\end{split}
\end{equation}
The term proportional to $h^{IJK}$ in (\ref{Deformedcomponentaction})
is exactly the term we would expect to arise in the twistor string
dual to the self--dual part of the marginally deformed theory. 
To obtain a more geometric understanding of the deformation, we 
notice that we can account for the additional term by introducing 
non--anticommutativity
between the three $\psi^I$ coordinates of $\Supertwistor$, while keeping 
the $\xi$ coordinate  anticommuting as before. Thus we
assume the following nontrivial non-anticommutation relation, with all 
other coordinates commuting or anticommuting as usual:
\begin{equation}
\{\psi^I,\psi^J\}=h^{IJQ}\gre_{QKL}\psi^K\psi^L
\end{equation}
and define a corresponding star product between superfields:
\begin{equation} \label{SDstarproduct}
\Acal\ast\Bcal= 
\Acal\Bcal+\Acal\overleftarrow{\frac{\p}{\p \psi^{I}}}
\frac{1}{2} h^{IJQ}\epsilon_{QKL}\psi^{K}\psi^{L}
\overrightarrow {\frac{\p}{\p \psi^{J}}}\Bcal.   
\end{equation}
It can be shown that, despite the coordinate dependence of the 
non--anticommutativity parameter,  this star product is associative 
when $h^{IJK}$ is totally symmetric, as we have assumed to be the case.
Also, exactly because it depends on the fermionic coordinates, all 
higher order terms vanish so the star product is exact. Equivalently,
the term in Fig. 1 of order $\Ocal(V^2)$ vanishes when $V$ is of the form
(\ref{VV}), as do all higher order terms.

Using the star product, we can now rewrite our deformation of holomorphic
Chern--Simons in terms of the superfield $\Acal$ in a very simple way:
\begin{equation}
\Scal_{def}=\half\int_{D5}\Omega\wedge
\Tr\left(\Acal\ast\dbar \Acal+\frac{2}{3}\Acal\ast\Acal\ast\Acal\right).
\end{equation}
The $\ast$ is now defined to also encompass the usual open string field 
theory star product, which in this case is just $\wedge$. Expanding this
action in components according to (\ref{Superfield}) (where the $\psi$s within each 
superfield are normal ordered because of the $\gre_{IJK}$s that appear, and thus we
need not introduce star products among them)
and integrating over $\xi,\psi^I$ we recover the
component action (\ref{Deformedcomponentaction}).

To summarise the discussion so far, we have found how to deform the holomorphic
Chern--Simons action to account for the extra vertex in the 
self--dual Leigh--Strassler
theory relative to self--dual $\Ncal=4$ SYM. The deformation involved turning
on a closed string background in the fermionic directions of $\Supertwistor$, 
which we interpreted as a non--anticommutative deformation. 
 
However our original goal was to calculate amplitudes in the full,
\emph{non--self dual} Leigh--Strassler theories, and it is easy to see that
to achieve that we need to do something more. First, looking at the amplitude
prescription (\ref{Amplitudeformula}) we see that we will have to decide how 
to multiply wavefunctions at different points. So we will have to generalise
(\ref{SDstarproduct}) to account for different coordinates $\psi^I_1,\psi^I_2$ etc.
More importantly, the full Leigh--Strassler
action (\ref{LSaction}) contains also the conjugate tensor 
$\hb_{IJK}$, which has not 
appeared on the string side so far. Since the amplitude prescription
was motivated by considerations of $D1$--branes, understanding the appearance 
of $\hb_{IJK}$ at a fundamental
level would probably require a better understanding of $D1$--brane 
interactions with the background than is available at the moment. 
So we will take
a practical approach, and simply modify the star product (\ref{SDstarproduct})
to incorporate $\hb_{IJK}$ in the most obvious way. We thus define 
\begin{equation} \label{Starproduct}
f(\psi_1)\ast g(\psi_2)=f(\psi_1)g(\psi_2)+\half \Vcal^{IJ}_{\;\;\;KL}
\left(f(\psi_1)\overleftarrow{\frac{\p}{\p\psi^I_1}}\right)\psi^K_1
\psi^L_2\left(\overrightarrow{\frac{\p}{\p\psi^J_2}}g(\psi_2)\right)\;+\cdots
\end{equation}
where $\psi_1$ and $\psi_2$ are the fermionic coordinates of two 
different wavefunctions, and 
the tensor $\Vcal^{IJ}_{\;\;\;KL}$ is defined to be
\begin{equation} \label{htensor}
\Vcal^{IJ}_{\;\;\;KL}=h^{IJQ}\gre_{QKL}+\gre^{IJQ} \hb_{QKL}\;.
\end{equation}

Using this new star product, we can generalise the prescription 
(\ref{Amplitudeformula}) for calculating amplitudes in the obvious way
(we now ignore the bosonic integration which is exactly as before):
\begin{equation} \label{DeformedFormula}
\Acal_{(n)}=\int\diff^8\theta\, w_1\ast w_2\ast\cdots\ast w_n 
\frac{1}{\ip{12}\ip{23}\cdots\ip{n1}}\;.
\end{equation}

We will now show, through a few simple examples, that this modification
of Witten's prescription does indeed generate amplitudes in the
marginally deformed theories. It is crucial to note, however, that
the new product (\ref{Starproduct}) is no longer associative starting
at second order in the deformation parameters, and thus the amplitudes
we calculate are only expected to match with field theory ones at
linear order. Determining the higher orders in the star product 
in order to regain associativity should allow us to calculate amplitudes
beyond linear order. This is currently work in progress.

As a first example we can take the four--gluon MHV amplitude $A_1A_2G_3G_4$, 
with $w(A)=1$ and $w(G)=1/3!\xi\psi^I\psi^J\psi^K\gre_{IJK}$ (see (\ref{Superfield})). 
Then the only nontrivial star product in (\ref{DeformedFormula}) is between 
$w_3$ and $w_4$, of the form $(\psi_3\psi_3\psi_3)\ast(\psi_4\psi_4\psi_4)$. Calculating
this product and integrating over the fermionic coordinates (using the fermionic part 
of the embedding equation (\ref{Embedding}) to relate the $\psi$s with the $\theta$ 
coordinates that we need to integrate over)  leaves us with the result
\begin{equation}
\Acal_{(4)}^{AAGG}=\frac{1}{3!3!}\left[\gre^{IJK}\gre^{MNP}
+\frac{9}{2}\Vcal^{KM}_{\;\;\;XY}\gre^{IJX}\gre^{YNP}\right]
\gre_{IJK}\gre_{MNP}\frac{\ip{34}^4}{\ip{12}\ip{23}\ip{34}\ip{41}}\;.
\end{equation}
It might seem that this amplitude will have a contribution at $\Ocal(\Vcal)$,
but working out the various contractions we find that it will actually have to be of 
the form  $\Vcal^{KM}_{\;\;\;KM}$ which
vanishes because of the symmetry properties of $\Vcal$ (\ref{htensor}). 
We conclude that the all--gluon amplitude is independent of the deformation which
matches our field theory expectations since (at tree level) this amplitude 
contains no vertices coming from the superpotential and thus is the
same in $\Ncal=4$ SYM and the deformed theory.

As our second and final example we take the amplitude $(\chi_I\chi_J\tchi^K\tchi^L)$
which will clearly be affected by the deformation. This amplitude is
also interesting because it is the sum of two Feynman diagrams:
\begin{figure}[ht] \label{Diagrams}
\begin{center}
\begin{picture}(300,50)(0,0)
\put(-30,0){
\Vertex(50,25){2} \Vertex(100,25){2}
\SetColor{BrickRed}
\DashLine(50,25)(100,25){1}
\SetColor{Blue}
\Line(25,10)(50,25)
\Line(25,40)(50,25)
\Line(100,25)(125,40)
\Line(100,25)(125,10)
\Text(20,40)[r]{$\chi_{I,1}$}\Text(20,10)[r]{$\chi_{J,2}$}
\Text(130,10)[l]{$\tchi^K_3$}\Text(130,40)[l]{$\tchi^L_4$}
\Text(60,27)[b]{$\phi_Q$}\Text(90,27)[b]{$\tphi^Q$}
\Text(75,1)[c]{\bf{(a)}}
}
\put(170,0){
\Vertex(50,25){2} \Vertex(100,25){2}
\SetColor{BrickRed}
\Photon(50,25)(100,25){2}{5}
\SetColor{Blue}
\Line(25,10)(50,25)
\Line(25,40)(50,25)
\Line(100,25)(125,40)
\Line(100,25)(125,10)
\Text(20,40)[r]{$\tchi^L_4$}\Text(20,10)[r]{$\chi_{I,1}$}
\Text(130,10)[l]{$\chi_{J,2}$}\Text(130,40)[l]{$\tchi^K_3$}
\Text(60,29)[b]{$A$}\Text(90,29)[b]{$A$}
\Text(75,1)[c]{\bf{(b)}}
}
\end{picture}
\caption{The two Feynman diagrams that contribute to tree--level $(\chi\chi\tchi\tchi)$ scattering.}
\end{center}
\end{figure}
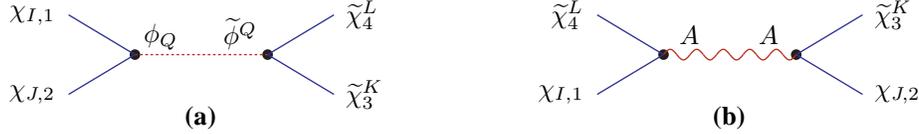

The first channel ((a) in Fig.2) is via scalar exchange, while the second one contains
only vertices involving gluons and thus should be unaffected by the
marginal deformation. Our amplitude prescription gives
\begin{equation}
\Acal_{(4)}^{\chi\chi\tchi\tchi}=\int\diff^8\theta\frac{1}{4}
\psi^I_1\ast \psi^J_2\ast(\xi_3\psi^M_3\psi^N_3)\ast(\xi_4\psi^P_4\psi^Q_4)\gre_{MNK}\gre_{PQL}
\frac{1}{\ip{12}\ip{23}\ip{34}\ip{41}}\;.
\end{equation}
Calculating the star product $\psi\ast\psi\ast(\psi\psi)\ast(\psi\psi)$
and integrating over superspace, we find that the answer at $\Ocal(V)$ 
is a sum of two terms:
\begin{equation} \label{chifourdeformed}
\Acal_{(4)}^{\chi\chi\tchi\tchi}=
-\left[\gre^{IJQ}\gre_{QKL}+h^{IJQ}\gre_{QKL}
+\gre^{IJQ}\hb_{QKL}\right]\frac{\ip{34}}{\ip{12}}
-\delta^I_{\;\;L}\delta^{J}_{\;\;K}\frac{\ip{34}^2}{\ip{23}\ip{41}}\;.
\end{equation}
Inspection of the momentum factors reveals that the term with the
$h,\hb$ dependence indeed corresponds to the first Feynman diagram in Fig.2, 
while the other term (corresponding to case (b) in Fig.2) does not receive any
corrections, in line with our field theory expectations.

More examples of low--point amplitudes can be found in 
\cite{KulaxiziZoubos04} and all agree with Feynman diagram calculations
in the Leigh--Strassler theories. Note also that since positive 
helicity gluons have no $\psi$ dependence, we can add any number of
them to obtain an n--point amplitude without introducing new star 
products. Furthermore, although we have exclusively discussed MHV amplitudes, we
do not expect any difficulty in extending (\ref{DeformedFormula}) 
to the non--MHV case.

In conclusion, we have shown that the twistor string formalism can
very naturally describe the Leigh--Strassler deformations of the
$\Ncal=4$ theory simply by considering the B--model on a 
non--anticommutative version of supertwistor space $\Supertwistor$. 
There are several aspects that need to be explored further, with
the main open issue being  understanding the higher orders
in the star product (\ref{Starproduct}).
This understanding will hopefully lead to a full (i.e. not only at
first order in the deformation parameters) reformulation of the marginal
deformations of $\Ncal=4$ SYM at tree level as a twistor string theory. 

Assuming this, our results \cite{KulaxiziZoubos04}, 
combined with those of \cite{ParkRey04,Giombietal04} which considered 
the twistor string dual of superconformal quiver gauge theories, give us
confidence that the B--model formalism is powerful enough to potentially describe 
any finite four--dimensional gauge theory. However, one still needs to develop more intuition
on how to extend the formalism to non--conformal cases (for recent progress in 
that direction, see \cite{Chiouetal05a,Chiouetal05b}), which would be 
necessary to obtain a more precise understanding of, for instance, the CSW 
method from twistor strings than what is available at the moment. We believe that
the work reviewed here is a step towards that goal.

\begin{acknowledgement}
 The second author would like to thank the organisers
of the RTN meeting ``Constituents, Fundamental Forces and Symmetries of
the Universe'' held in Corfu, Greece, September 20-26, 2005,
for giving him the opportunity to present our work there.
\end{acknowledgement}

\end{document}